\documentstyle[aps,epsf]{revtex}
\draft
\title{Cluster formulation of spin glasses and the frustrated percolation
model:\\ statics and dynamics}
\author{Antonio de Candia, Vittorio Cataudella, Antonio Coniglio}
\address{Dipartimento di Scienze Fisiche,
         Mostra d'Oltremare, pad. 19, 80125 Napoli, Italy}
\address{INFM, Sezione di Napoli, Napoli, Italy}
\begin{document}
\twocolumn
\wideabs{%
\maketitle
\begin{abstract}
We study the properties of the $q$-state frustrated bond percolation model
by a Monte Carlo ``bond flip'' dynamics, using an algorithm originally 
devised by Sweeny and suitably modified to treat the presence of frustration.
For $q=2$ the model gives the cluster formulation of the Edwards Anderson
spin glass.
We analyze the percolation transition of the model, and find that it falls 
in the universality class of the $q/2$-state ferromagnetic Potts model.
We then investigate the bond flip dynamics of the model, 
and find that, while for temperatures higher than the percolation
transition $T_p$ the relaxation functions are fitted by a single
exponential, for $T<T_p$ they show a two step decay, reminiscent of 
the relaxation of glass forming liquids.
The long time decay ($\alpha$-relaxation) is well fitted for $T<T_p$
by a stretched exponential function, showing that in this
model the relevant mechanism for the appearing of stretched exponentials
is the percolation transition.
At very low temperatures the relaxation functions develop a long
{\em plateau}, as observed in glass forming liquids.
\end{abstract}
}
%
%
\section{Introduction}
Cluster concepts have been extremely useful in critical phenomena to 
elucidate the mechanism underlying a thermodynamical transition,
by providing a geometrical interpretation of thermodynamic correlations. In
the Ising model defined by the Hamiltonian
\begin{equation}
{\cal H}=-J\sum_{\langle{ij}\rangle}S_i S_j,
\label{eq_ising}
\end{equation}
a correct definition of clusters was given by Kasteleyn and Fortuin 
\cite{ref_kf} and by Coniglio and Klein \cite{ref_ck}. In the 
Coniglio-Klein approach, one throws a bond between pairs of parallel 
nearest neighbor 
spins with a probability $p=1-e^{-2\beta J}$,
where $\beta=1/k_BT$.
By summing over the spin configurations with the Boltzmann weight,
the partition function of the model defined by Eq.
(\ref{eq_ising}) can be  written as a sum over bond configurations $C$,
\begin{equation}
Z=\sum_C e^{\mu b(C)}\,q^{N(C)},
\label{eq_ising_Z}
\end{equation}
where $q=2$, $b(C)$ is the number of bonds in the configuration $C$, 
$N(C)$ the number of connected clusters, and 
$\mu=\log(\frac{p}{1-p})=\log(e^{2\beta J}-1)$
is the chemical potential of the bonds.
Thermodynamic averages
can be related in this approach to corresponding percolative quantities.
One finds that the clusters represent spin fluctuations, and
percolate at the Ising critical temperature with Ising critical exponents.

This approach can be extended to the Potts model, in which spins can have
$q\neq 2$ different states.
In this case the parameter $q$ in Eq. (\ref{eq_ising_Z})
can assume a value $q\neq 2$, and for every value of $q$ the percolation
model has the same critical temperature and exponents of the corresponding
Potts model. For $q=1$ one recovers the random bond percolation.
Sweeny \cite{ref_sweeny} studied the weighted percolation problem defined
by Eq. (\ref{eq_ising_Z})  by Monte Carlo techniques  
on a two-dimensional square lattice. 
He showed that a simulation based on this approach does not 
suffer from
critical slowing down for $q<4$. In a few Monte Carlo steps one can 
equilibrate even a very large lattice at the critical temperature. Thus he
extracted informations about the critical point of the Potts model 
by measuring geometric quantities like
the mean cluster size at the transition point.
Later, the cluster approach was further elaborated 
by Swendsen and Wang \cite{ref_swang}
and Wolff \cite{ref_wolff}, by implementing an efficient cluster dynamics
to simulate directly the spin
model (\ref{eq_ising}), which drastically reduces
the critical slowing down of conventional spin flip simulations.

The cluster approach has been extended also to frustrated systems 
like spin glasses \cite{ref_cdilib,ref_varenna}. 
With respect to the ferromagnetic case,
new features appear, due to the phenomenon of frustration. In fact
the percolation model that one obtains has the same complexity of the 
original spin model, and is not useful to define fast Monte Carlo 
dynamics, as in the unfrustrated case. Nevertheless, it represents an
interesting tool to investigate the properties of 
frustrated spin systems from the geometrical point of view. Moreover, 
it can be considered on its own as a model of percolation in a frustrated
medium, that makes it of relevance in the study of systems in which 
frustration and connectivity play a central role, as the structural glasses.

To illustrate the cluster approach to frustrated spin models, let us consider
the Ising spin glass, defined by the Hamiltonian 
\begin{equation}
{\cal H}=-J\sum_{\langle ij\rangle} \epsilon_{ij}S_i S_j,
\label{eq_spinglass}
\end{equation}
where $\epsilon_{ij}$ are quenched random variables that can have the values
$\epsilon_{ij}=\pm 1$.
As in the ferromagnetic case, one throws a bond between nearest neighbor
spins that satisfy the interaction (in this case if
$\epsilon_{ij}S_i S_j=1$) with probability $p=1-e^{-2\beta J}$.
The crucial difference with the ferromagnetic case is that
in general not all the interactions can be satisfied simultaneously.
Indeed a closed loop  such that the product of 
$\epsilon_{ij}$ along the loop is negative does not admit any spin 
configuration satisfying all the interactions, and is called frustrated.
Since bonds can be put only between spins that satisfy the interaction, 
one cannot put bonds on the lattice that close a frustrated loop. In terms
of bond configurations, the partition function of the Ising spin glass 
turns out to be
\begin{equation}
Z={\sum_C}^\ast e^{\mu b(C)}\,q^{N(C)},
\label{eq_sg_Z}
\end{equation} 
where $q=2$, $\mu=\log(\frac{p}{1-p})=\log(e^{2\beta J}-1)$,
$b(C)$ is the number of bonds, and $N(C)$ the number of clusters
in the configuration $C$.
Here the sum $\sum_C^\ast$ is extended to all the bond configurations that
do not contain a frustrated loop. 
Therefore, in this cluster formalism, the only difference between the 
Ising spin glass and the ferromagnetic model is that, in the ferromagnetic
case (\ref{eq_ising_Z}) the sum is over all the bond configurations, 
while in the spin glass (\ref{eq_sg_Z}) due to the geometrical
constraint the sum is restricted to a subset of the bond configurations. 
In particular the ground state at $T=0$ ($\mu=\infty$)
is obtained by maximizing the number of bonds, under the constraint that
the bond configuration does not contain a frustrated loop.

Due to frustration, clusters defined in the spin glass model
no longer correspond to thermodynamical fluctuations.
In fact the correlations between spins can be either positive (if they 
propagate along a path that contains an even number of negative 
interactions) or negative (if the path contains an odd number of negative
interactions), so they interphere and are canceled out at least in part
\cite{ref_cdilib,ref_varenna}.
On the other hand connectivity is always positive and, as a result,
clusters in the spin glass model percolate at a higher
temperature $T_p$ respect to the critical temperature $T_{SG}$. In the 
three dimensional Ising spin glass simulations show that 
$T_p\simeq 3.95 J/k_B$
\cite{ref_cdearc}, while $T_{SG}\simeq 1.11 J/k_B$ \cite{ref_young}.

As in the ferromagnetic case, the model defined by Eq. (\ref{eq_sg_Z}) 
can be extended
to values $q\neq 2$ of the spin multiplicity. 
We call this model the ``$q$-state frustrated percolation model''.
For $q$ integer and even, Eq. (\ref{eq_sg_Z}) is the partition function
of the model defined by the Hamiltonian \cite{ref_vittorio}
\begin{equation}
{\cal H}=-J\sum_{\langle ij\rangle} 
\delta_{\sigma_i\sigma_j} (\epsilon_{ij}S_i S_j+1),
\label{eq_potts}
\end{equation}
where $\sigma_i=1\ldots q/2$ are Potts spins, and 
$\delta_{\sigma_i\sigma_j}$ is the Kroneker's delta.
In analogy to the case $q=2$, the model is 
expected to have two transitions, one at a temperature 
$T_{SG}(q)$ corresponding to freezing of Ising spins, and the other at
a temperature $T_p(q)>T_{SG}(q)$ corresponding to the percolation of
the clusters. A renormalization group calculation carried over a hierarchical
lattice \cite{ref_pezzella} has confirmed this expectation, and has
shown that the transition at $T_{SG}$ should be in the universality
class of the Ising spin glass, no matter what is the value of $q$, while
the percolation transition at $T_p$ should be in the universality class 
of the $q/2$ state Potts model. 
For $q=1$ the model assume a very simple form,
as the factor $q^{N(C)}$ disappears. The resulting model has been called
``frustrated percolation'', and can be viewed as a simple  model of 
percolation in a frustrated medium. Despite its simplicity,
its dynamical properties  exhibit a complex behavior, with features 
in common with both structural glasses and spin glasses.

In this paper, we perform a Monte Carlo study of the 
$q$-state frustrated percolation model on a two-dimensional square lattice,
using a ``bond flip'' dynamics, in which 
bonds are added and removed from the lattice with appropriate probabilities,
in order to satisfy the principle of detailed balance. To do this, we have
realized an algorithm that allows to determine the connectedness of two given
sites, and the presence of frustrated loops, with a time that in the worst 
case (at the percolation temperature) scales only with the logarithm of the
lattice size, as we describe in Sect. \ref{sec_algorithm}. The algorithm is
a modification of the algorithm used by Sweeny in the ferromagnetic case
\cite{ref_sweeny}.
In Sect. \ref{sec_percola} we report our results on the percolation 
transition of the model for $q=1,2,4$. Note that while for $q$ multiple 
of $2$ the percolation transition can be studied also by conventional 
spin flip \cite{ref_fc}, 
for other values of $q$ the ``bond flip'' dynamics is the 
only way to simulate the model. 
In Sect. \ref{sec_stretched} and \ref{sec_dynamics} we study the 
dynamical properties of the model, analyzing the relaxation
functions of the number of bonds. 
\section{Monte Carlo algorithm}
\label{sec_algorithm}
We have implemented a Monte Carlo algorithm to simulate the bond percolation 
model defined by Eq. (\ref{eq_sg_Z}) on a two-dimensional square
lattice, which can be applied 
for any value of the parameter $q\in[0,\infty)$. 
The interactions $\epsilon_{ij}$ between pairs $\langle{ij}\rangle$
of nearest neighbor sites are set at the beginning to a value $+1$ or $-1$ 
randomly, with equal probability. These variables are quenched, and their 
state is not changed by the dynamics.

Each edge of the lattice, that is each pair of nearest neighbor sites, can be 
in two possible states: connected by a bond or not. At each step of the
dynamics, we try to flip the state of an edge chosen randomly, with a 
probability determined in such a way to satisfy the principle of detailed 
balance. If we try to remove a bond from the system, or to add a bond that do 
not close a frustrated loop, then the probability of carrying out the ``bond 
flip'' will be
\begin{equation}
P_{\text{flip}}=\min(1,e^{\mu\delta b}\,q^{\delta N}),
\end{equation}
where $\delta b$ is the change in the number of bonds, $\delta N$ is the 
change in the number of connected clusters. 
If we are trying to add a bond that closes a frustrated loop, 
then we have simply $P_{\text{flip}}=0$. 
A Monte Carlo Step (MCS) is 
defined as $2V$ single bond flip trials, where $V=L^2$ is the total 
number of sites, and  $2V$ the total number of edges of the lattice.

\begin{figure}
\begin{center}
\mbox{(a)\epsfysize=4cm\epsfbox{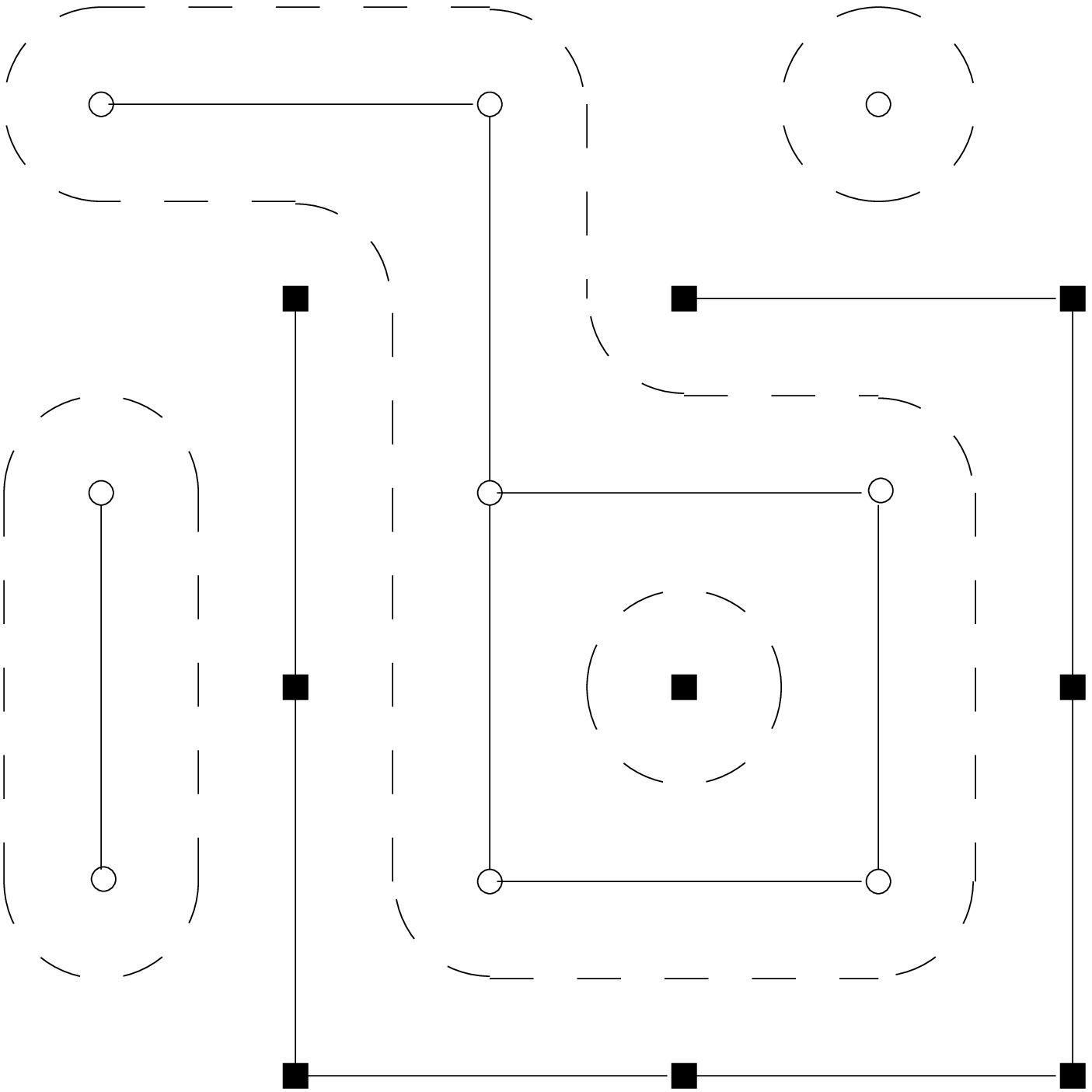}}
\end{center}
\begin{center}
\mbox{(b)\epsfysize=4cm\epsfbox{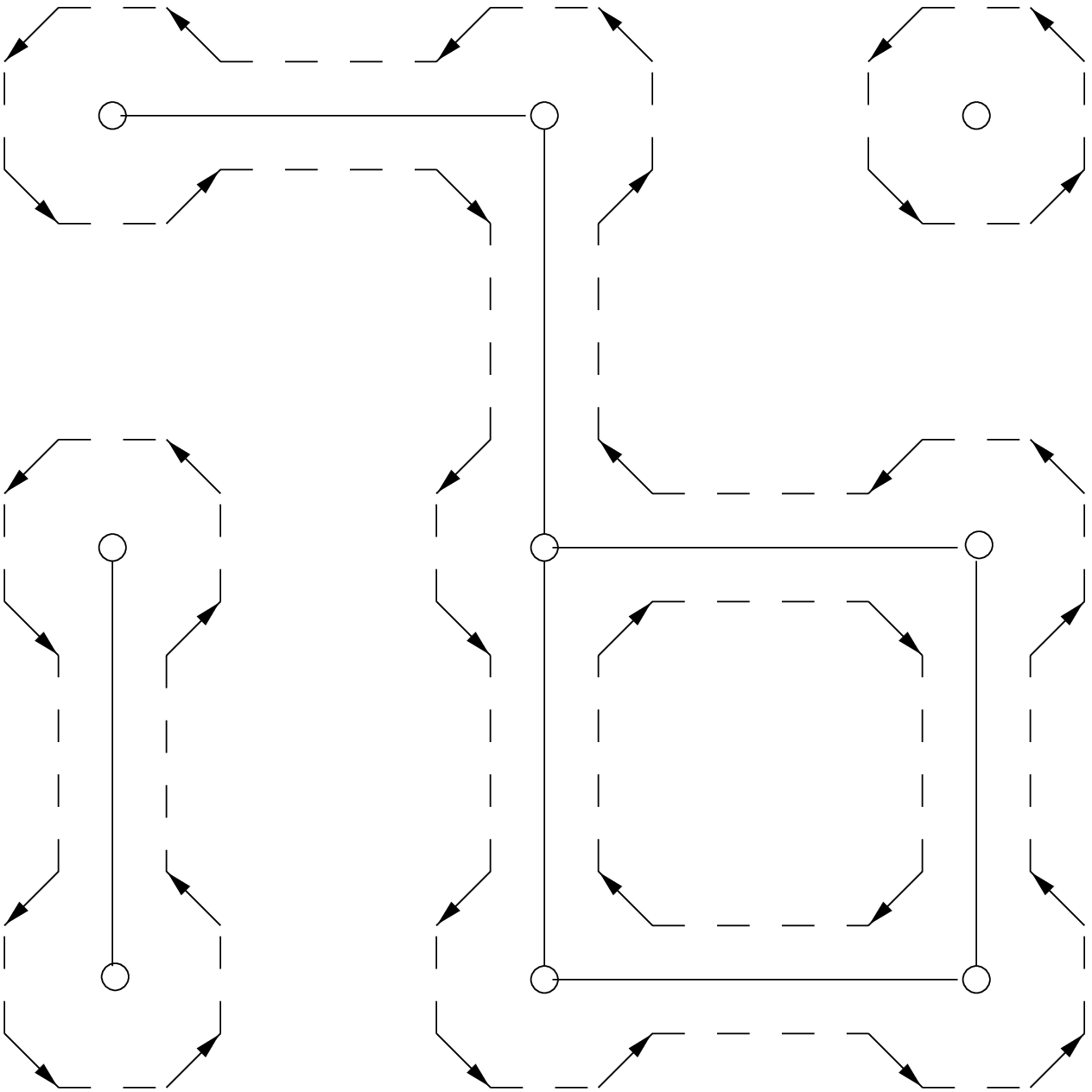}}
\end{center}
\bigskip
\caption{(a) Bonds, dual bonds, and chains. Open circles represent sites of 
the original lattice; solid squares sites of the dual lattice; solid lines
bonds and dual bonds; dashed lines represent chains. (b) Pointers forming 
chains on the lattice. Each pointer points to an ALONG pointer in the direction
of the arrow, and to an UP pointer in the opposite direction (not
necessarily the nearest ones).}
\label{fig_sw1}
\end{figure}

The nontrivial point here is to determine the change in number of connected 
clusters, and to verify if a bond added between two given sites closes or not 
a frustrated loop. 
To do this, we have used the algorithm used by Sweeny 
in the ferromagnetic case
\cite{ref_sweeny}, suitably modified to treat the frustration occurrence. 
Consider a two-dimensional square lattice, together with its dual lattice. 
If a bond is present on the lattice, then its dual bond is absent, and 
viceversa. The boundaries between connected clusters on the lattice and on its 
dual will form a collection of closed loops, as shown in 
Fig. \ref{fig_sw1}(a). These loops are represented in the computer as
chains of pointers. Each site on the lattice has four pointers adjacent to it, 
as shown in Fig. \ref{fig_sw1}(b). At the beginning of the simulation,
pointers are organized in a
hierarchical way, by giving them a defined ``level'',
that do not change in the following. A fraction $(4^{-n}-4^{-(n+1)})$ of the
pointers are at level $n=0,1,\ldots,n_{\text{max}}-1$, and
$4^{-n_{\text{max}}}$ at level $n_{\text{max}}$, where level
$n_{\text{max}}$ is chosen so that it counts not more than four pointers. 
Chains
then are formed by making each pointer point to other two pointers of the 
chain, one in the direction of the arrows, called the ALONG pointer, and one
in the opposite direction, called the UP pointer. The ALONG pointer must be 
at least at the same level of the one pointing to it, and the UP pointer at
a higher level (except if there are no higher level pointers
in the chain), so they in general do not correspond to the nearest pointers
in the chain.

\begin{figure}
\begin{center}
\mbox{(a)\epsfysize=2.5cm\epsfbox{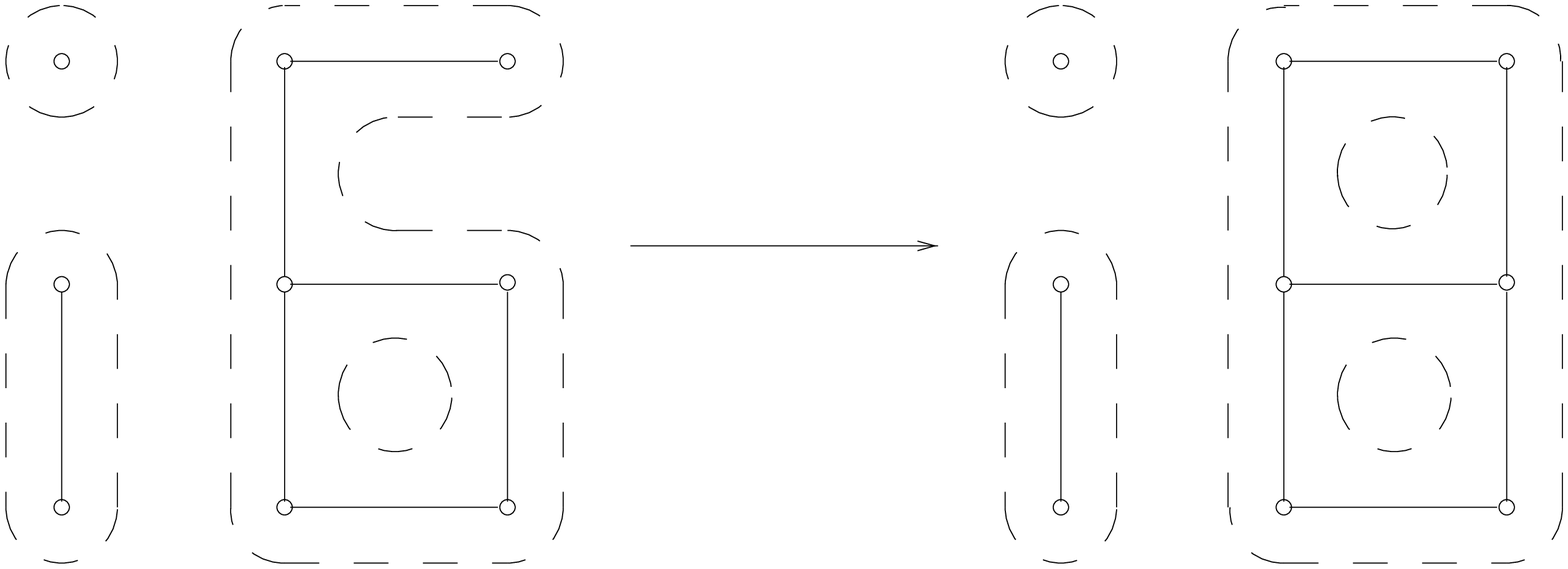}}
\end{center}
\bigskip
\begin{center}
\mbox{(b)\epsfysize=2.5cm\epsfbox{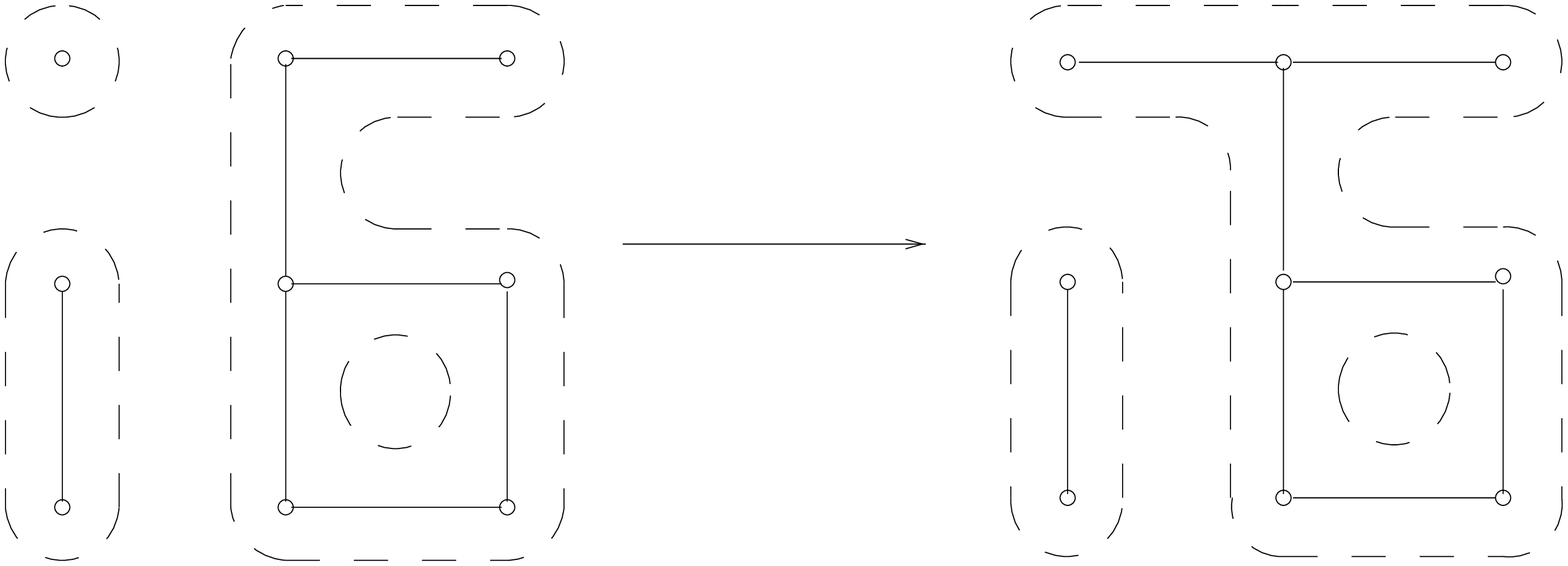}}
\end{center}
\bigskip
\caption{Addition of a bond to the lattice: (a) the bond links two sites 
already belonging to the same cluster, and cuts a single chain into two 
distinct chains; (b) the bond links two disconnected clusters, and joins two 
distinct chains into a single chain.} 
\label{fig_sw2}
\end{figure}

When we add a bond to the lattice, (and remove its dual),
two things can happen: the bond links two sites already belonging to the same 
connected cluster, and therefore $\delta N=0$; the bond links two previously 
disconnected clusters, and $\delta N=-1$. In the first case the bond will cut 
a single chain into two distinct chains, see Fig. \ref{fig_sw2}(a),
while in the second case it will join 
two distinct chains into a single chain, see Fig. \ref{fig_sw2}(b).
When we remove a bond from the lattice, (and add its dual), it happens the 
other way round.
Using this auxiliary data structure, one can determine if two 
given chain pointers belong to the same chain or not, and cut and rejoin 
chain segments, in a CPU time that grows only with the logarithm of the 
chain length.

\begin{figure}
\begin{center}
\mbox{\epsfysize=5cm\epsfbox{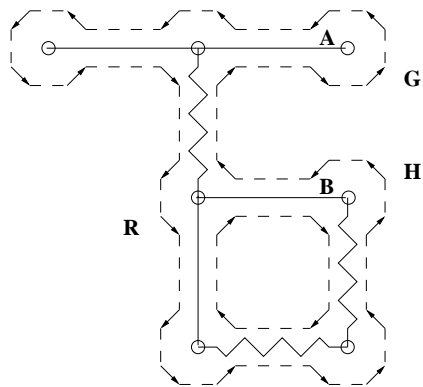}}
\end{center}
\bigskip
\caption{Determination of the number of antiferromagnetic bonds
along a path from site A to site B. Straight lines 
represent ferromagnetic bonds,
wavy lines antiferromagnetic ones.
The number of antiferromagnetic 
bonds skirted traversing the chain in the UP direction (opposite to the arrows)
from the pointer G adjacent to A
to the reference pointer R is 3, while from
H to R is 2. Making the difference, we find the number we seek,
that is 1.} 
\label{fig_sw3}
\end{figure}

To simulate the frustrated percolation model, 
we must also determine if a bond added to the
lattice closes a frustrated loop or not. To do this, we must be able to count 
the number of antiferromagnetic bonds encountered along a path that joins two 
given sites A and B. 
This can be done if every chain pointer contains information 
about the number of antiferromagnetic bonds ``skirted'' when one traverses 
the chain to its UP pointer.
Call this number the ``phase'' of the pointer respect to its UP pointer. 
We then go on jumping from each pointer to its UP pointer, and adding the 
relative phases, until we reach a reference pointer R in the chain.
Then the number we seek for is found as the difference between the phases
of two pointers adjacent to the sites A and B, respect to the reference 
pointer R, as shown in Fig. \ref{fig_sw3}. 
This reference pointer is chosen between the highest level pointers in the
chain, and is the UP pointer of all the pointers belonging to the highest
level in the chain.
When we cut and rejoin chain segments, 
we must update coherently the relative phases of the chain pointers involved,
and assure that there is one and only one reference pointer per chain.
\section{The percolation transition}
\label{sec_percola}
We have studied the percolation transition of the model defined by Eq.
(\ref{eq_sg_Z}), for $q=1,2,4$. For each value of $q$ we have simulated the
model for lattice sizes $L=32,64,128$. The histogram method
\cite{ref_histo} was used to analyze the data. 
In this Section, we use the probability $p$ as the
independent variable. It is connected to the temperature via the simple
relation $p=1-e^{-2\beta J}$.
For each value of $q$ and $L$,
sixteen probabilities were simulated around the percolation transition
point, taking $10^3$ MCS for thermalization, and between $10^4$ MCS
(for $L=128$) and $10^5$ MCS (for $L=32$) for the acquisition of the
histograms. Histograms were taken of the number of bonds; of the mean cluster
size, defined as $\displaystyle\frac{1}{V}\sum_s s^2 n_s$,
where $n_s$ is the number of clusters of
size $s$; and of the occurrence of a spanning cluster, defined 
as a cluster that spans from the bottom to the top of the lattice.
These histograms were used to calculate the average 
number of bonds $\langle{b}\rangle$,
the fluctuation in the number of bonds
$\langle{b^2}\rangle-\langle{b}\rangle^2$, the average 
mean cluster size $\chi$,
and the spanning probability $P_\infty$, in a whole interval of
probabilities around the percolation transition probability.

\begin{figure}
\begin{center}
\mbox{(a)\epsfysize=6cm\epsfbox{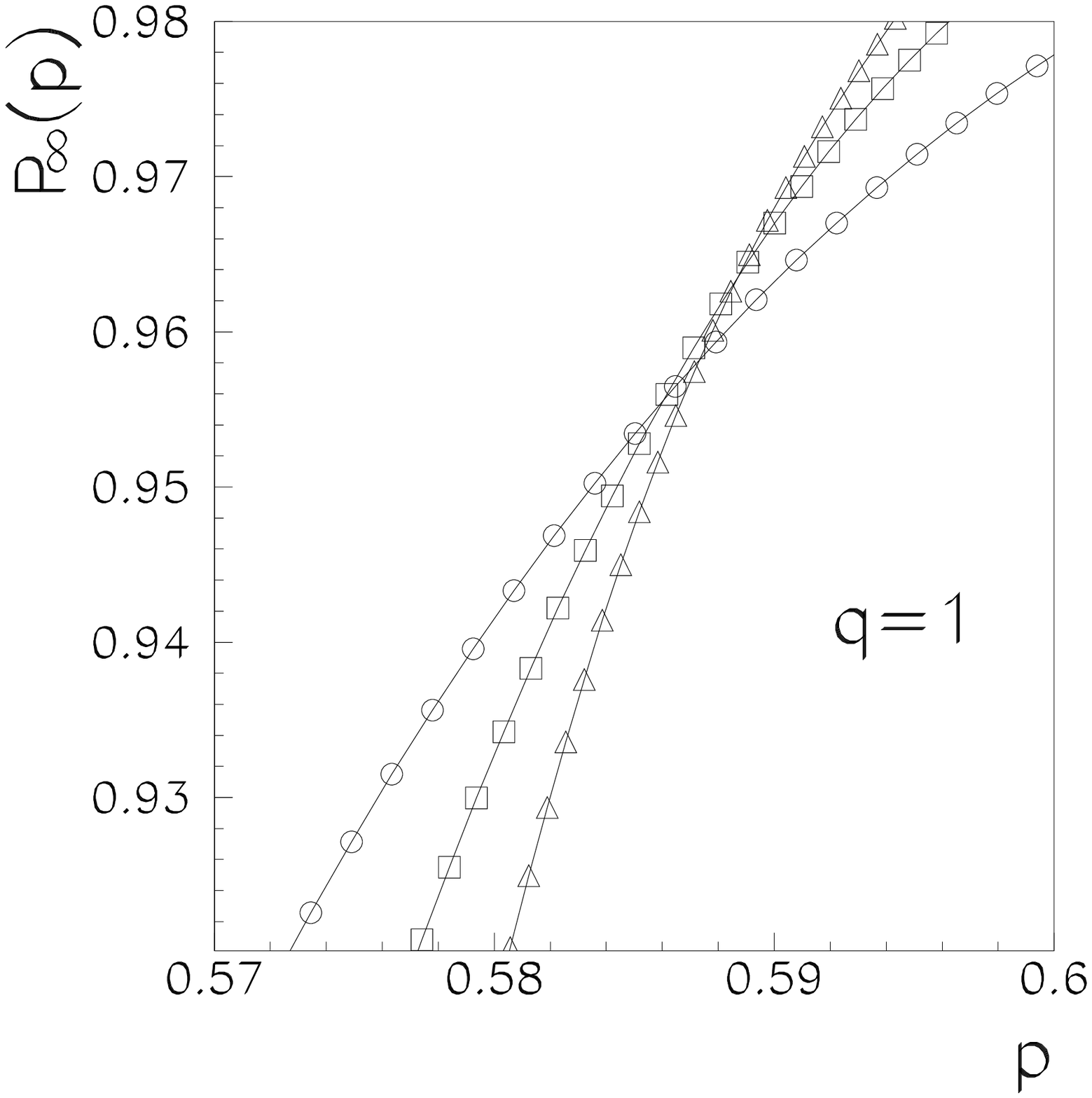}}
\end{center}
\bigskip
\begin{center}
\mbox{(b)\epsfysize=6cm\epsfbox{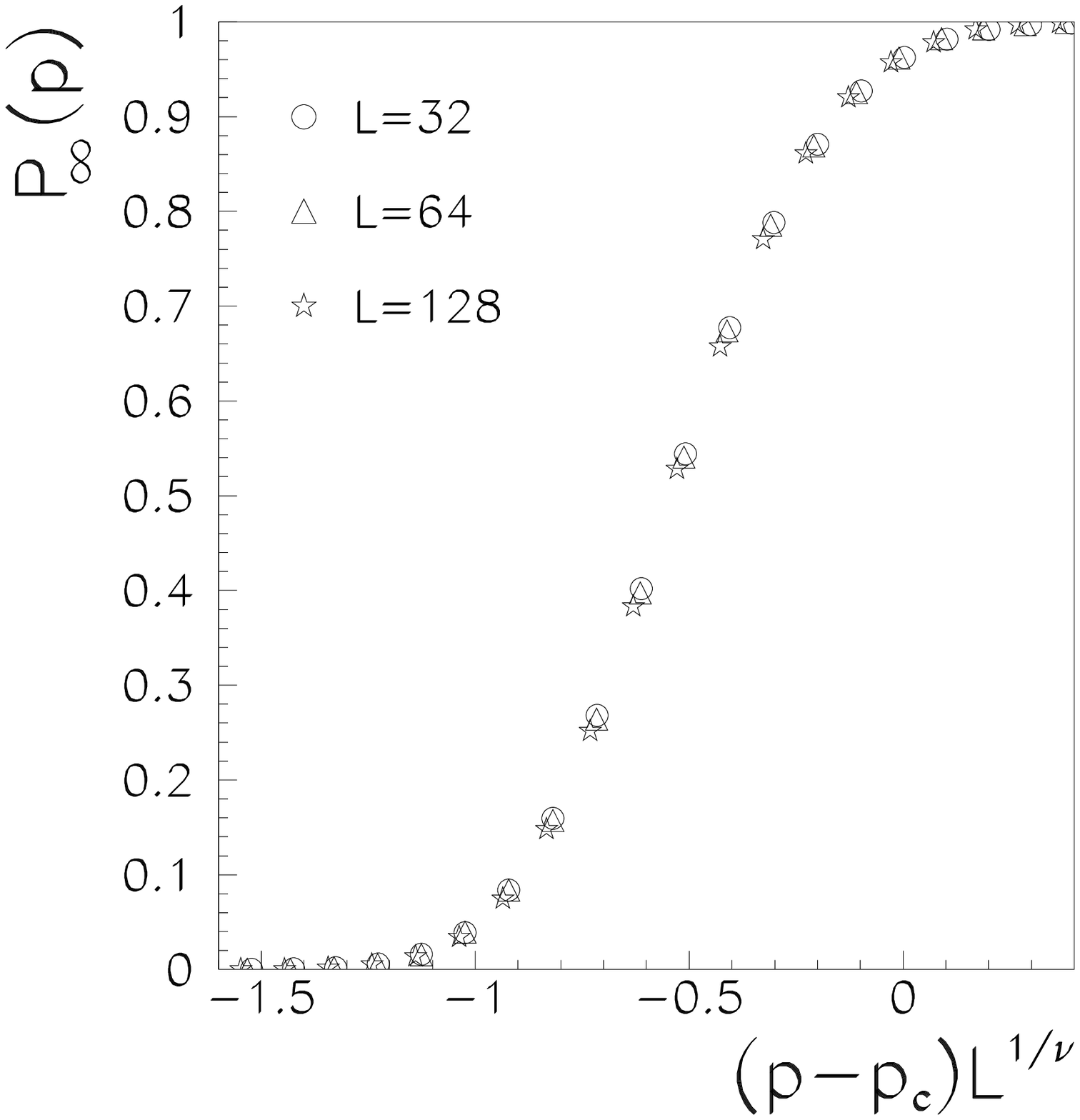}}
\end{center}
\bigskip
\caption{Measure of percolative quantities on the model with
$q=1$ and lattice sizes $L=32,64,128$. (a) Spanning probability $P_\infty(p)$
as a function of probability $p$.
(b) $P_\infty(p)$ as a function of $(p-p_c)L^{1/\nu}$,
with $p_c=0.589$ and $1/\nu=0.56$.}
\label{fig_percola1}
\end{figure}

\begin{figure}
\begin{center}
\mbox{(a)\epsfysize=6cm\epsfbox{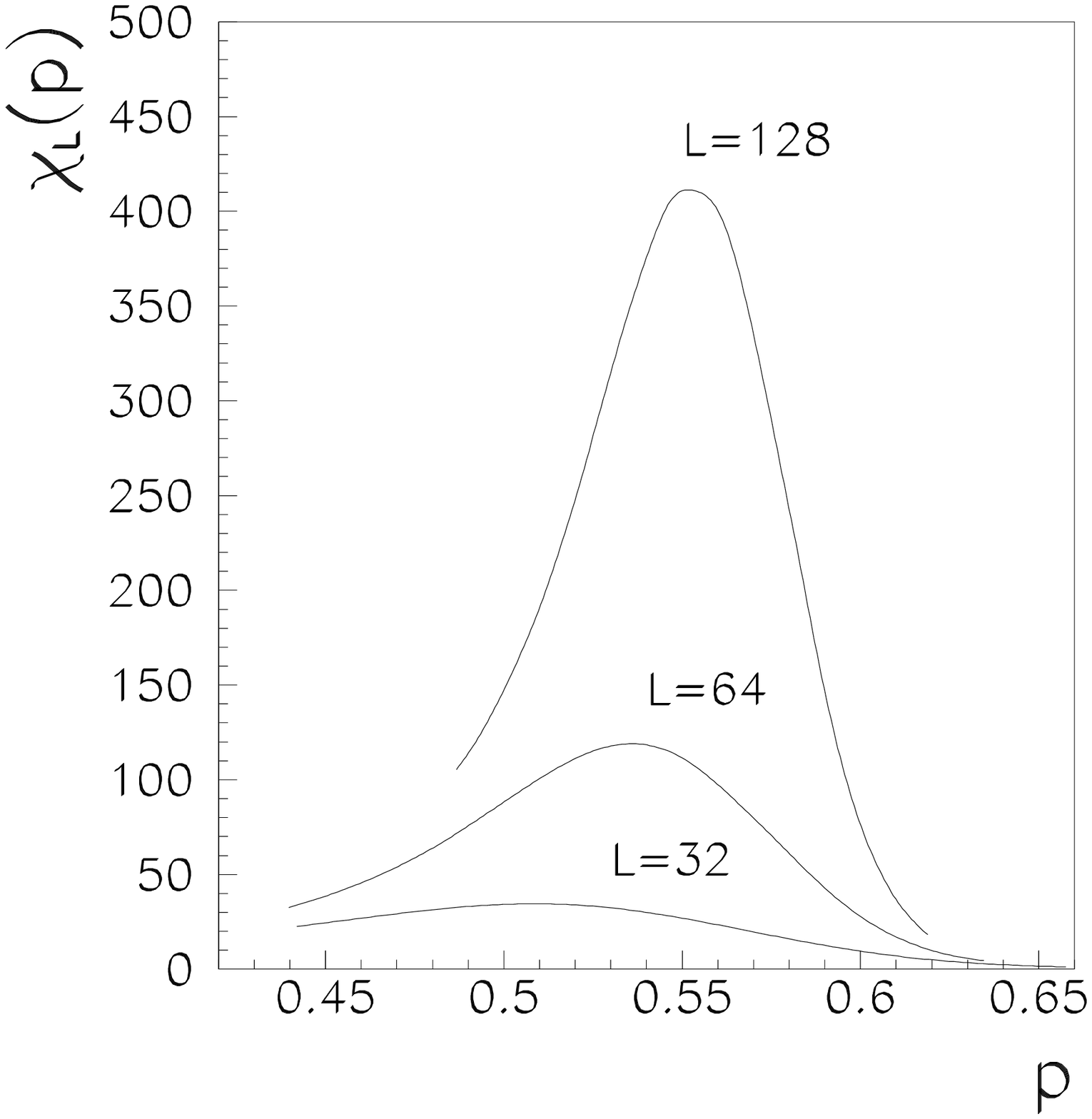}}
\end{center}
\bigskip
\begin{center}
\mbox{(b)\epsfysize=6cm\epsfbox{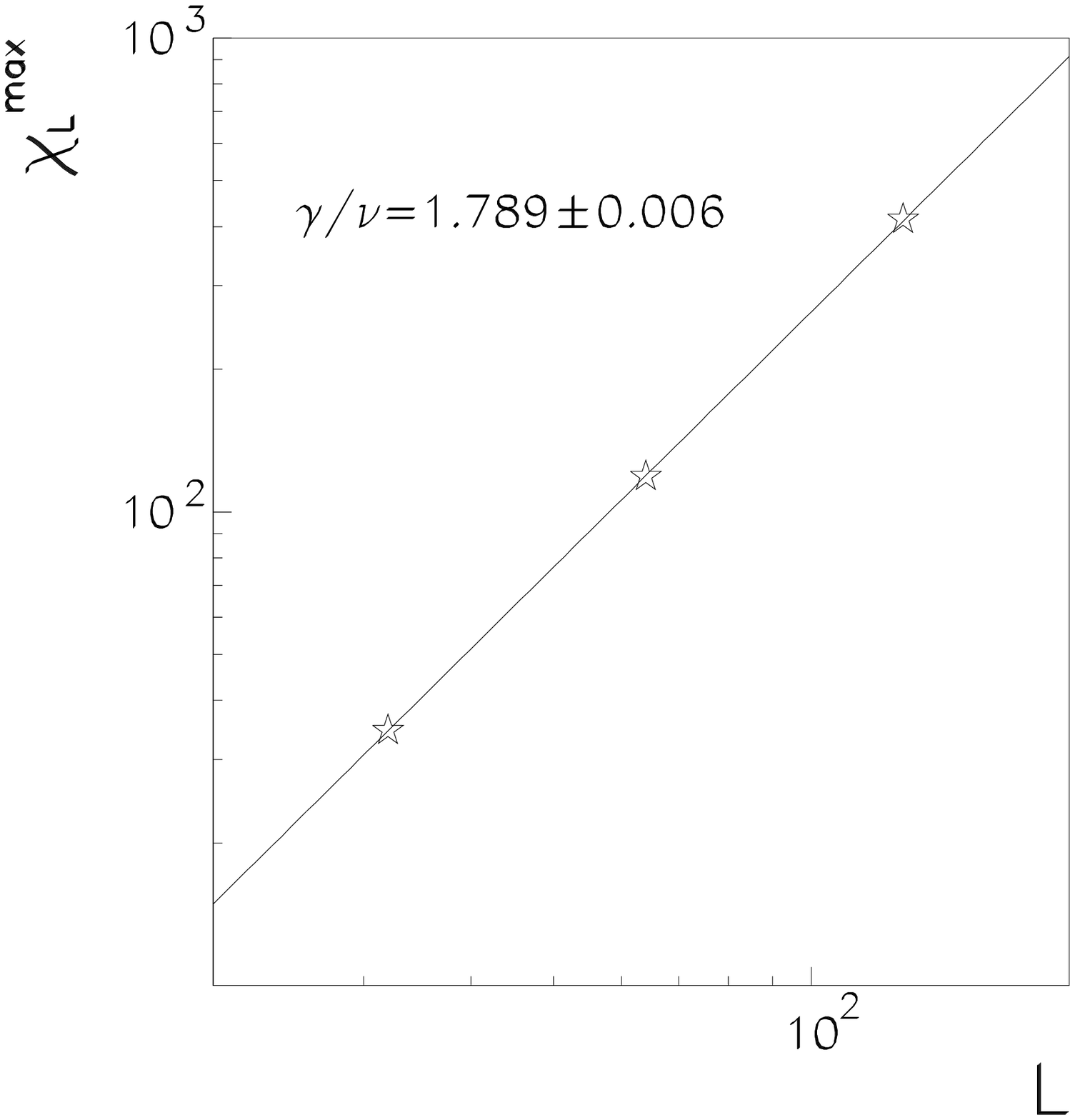}}
\end{center}
\bigskip
\caption{(a) Mean cluster size $\chi_L(p)$
as a function of probability $p$, for lattice sizes $L=32$, 64, 128.
(b) Maximum $\chi_L^{\text{max}}$ of $\chi_L(p)$ as a function of lattice
size $L$.}
\label{fig_percola2}
\end{figure}

The spanning probability $P_\infty(p)$ and the
mean cluster size $\chi(p)$, as a function of the lattice size $L$
and of the probability $p=1-e^{-2\beta J}$, around
the percolation probability $p_c$ should obey
the scaling laws \cite{ref_percola,ref_binder}
\begin{mathletters}
\begin{equation}
P_\infty(p) \simeq \widetilde{P}_\infty[L^{1/\nu}(p-p_c)],
\label{eq_Pscaling}
\end{equation}
\begin{equation}
\chi(p) \simeq  L^{\gamma/\nu}\widetilde{\chi}[L^{1/\nu}(p-p_c)],
\label{eq_chiscaling}
\end{equation}
\end{mathletters}
where $\gamma$ and $\nu$ are the critical exponents of mean cluster size
and connectivity length, $\widetilde{P}_\infty$ and $\widetilde{\chi}$ 
are universal functions.
Given Eq. (\ref{eq_Pscaling}), the value of the transition probability
$p_c$ can be evaluated from the point at which the curves $P_\infty(p)$ 
for different values of $L$ intersect. 
In Fig. \ref{fig_percola1}(a) are plotted
the measured curves $P_\infty(p)$ for $q=1$ and $L=32,64,128$.
One must extrapolate the value at which curves for $L,L^\prime\to\infty$
intersect. Then we have evaluated the critical exponent $1/\nu$, by choosing
the value that gives the best data collapse of the curves, when one plots
$P_\infty(p)$ in function of $L^{1/\nu}(p-p_c)$,
see Fig. \ref{fig_percola1}(b).
The errors on $p_c$ and $1/\nu$ were evaluated as the amplitudes of the
intervals within which a good data collapse was obtained.
Given Eq. (\ref{eq_chiscaling}), the mean cluster size $\chi(p)$ has a
maximum that scales as $L^{\gamma/\nu}$, see Fig. \ref{fig_percola2}(a). 
{From} a log-log plot of $\chi^{\text{max}}$ in function of $L$, 
we can extract the
exponent $\gamma/\nu$, making a linear fit, as shown in Fig. 
\ref{fig_percola2}(b).
Results are summarized in Tab. \ref{tab_critic}. These results are in good
agreement with the prediction that the percolation transition of the model
falls in the universality class of the $q/2$-state ferromagnetic Potts model
\cite{ref_pezzella}, 
whose critical exponents are reported in Tab. \ref{tab_potts} \cite{ref_wu}.

\begin{figure}
\begin{center}
\mbox{\epsfysize=6cm\epsfbox{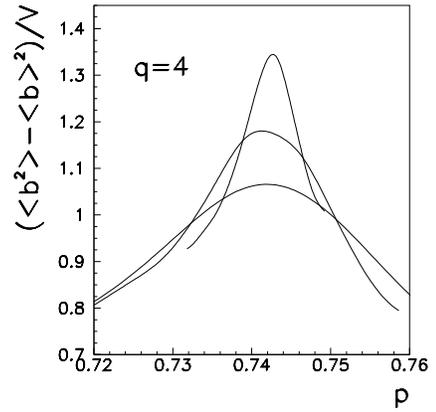}}
\end{center}
\bigskip
\caption{Fluctuation $\langle{b^2}\rangle-\langle{b}\rangle^2$ of the number
of bonds, divided by the number of sites V, as a function of $p$,
for $q=4$ and lattice sizes (from bottom to top) $L=32,64,128$.}
\label{fig_heat}
\end{figure}

Renormalization group calculations \cite{ref_pezzella} show that the model
should also have a singularity in the free energy density $F(p)$ at $p_c(q)$,
with a singular part 
$F_{\text{sing}}(p)\sim A(q) (p-p_c)^{2-\alpha(q/2)}$,
where $\alpha(q/2)$ is the specific heat
exponent of the $q/2$-state ferromagnetic Potts model. In particular
the model with $q=4$ should have a singularity corresponding to the
ferromagnetic Ising model, that is a logarithmic divergence of the
second derivative of the free energy (specific heat).
From Eq. (\ref{eq_sg_Z})
it is possible to show, that the derivatives of
$-\beta F=V^{-1}\log Z$
respect to $\mu$
are equal to the cumulants of the distribution of the number of bonds,
divided by the total number of sites $V$. As $\mu$ is  a regular function of
the probability $p$ for $0<p<1$, we conclude that the model with $q=4$ should
have a divergence in the second cumulant, that is the fluctuation,
of the number of bonds divided by $V$. For finite
size systems, we expect to see a peak whose maximum scales as $\log(L)$. In
Fig. \ref{fig_heat}  we show   our results for $q=4$ and $L=32,64,128$.
It is evident that there is a divergence, but statistical errors do not
allow to distinguish between logarithmic and a weak power law divergence.

For $q=2$ one would expect a singularity in the free energy,
at the percolation transition, characterized by an exponent
$2-\alpha(1)=8/3$, that is a divergence in the third cumulant of the
number of bonds. For the Ising spin glass ($q=2$)
this would imply a divergence in the third cumulant of the
distribution of the energy.
However several arguments, including a renormalization group 
calculation \cite{ref_pezzella}, 
predict that the singularity could be canceled out 
by the vanishing of the prefactor $A(q)$ for $q\to 2$.
This prediction could be checked by showing that the third
cumulant does not diverge, but
much more extensive simulations are needed to verify this prediction.
\section{Onset of stretched exponentials}
\label{sec_stretched}
We have studied the dynamical properties of the 
$q$-state frustrated percolation model for $q=1$, 2, and 4,
by calculating the autocorrelation function of the number of bonds.
This is defined as
\begin{equation}
F(t)=
\frac{\langle b(0)b(t)\rangle-\langle b\rangle^2}
{\langle b^2\rangle-\langle b\rangle^2},
\label{eq_relax}
\end{equation}
and is normalized so that $F(0)=1$, while for $t\to\infty$  it
relaxes to zero.
We simulated the model on a two-dimensional square lattice of size 
$32\times 32$, and took $10^5$ MCS for thermalization, and between
$10^5$ and $2\times 10^6$ MCS for acquisition. All the functions were 
then averaged over 16 different configurations of the interactions 
$\epsilon_{ij}$. Errors are evaluated as mean standard deviation 
of this last averaging.

\begin{figure}
\begin{center}
\mbox{(a)\epsfysize=7.5cm\epsfbox{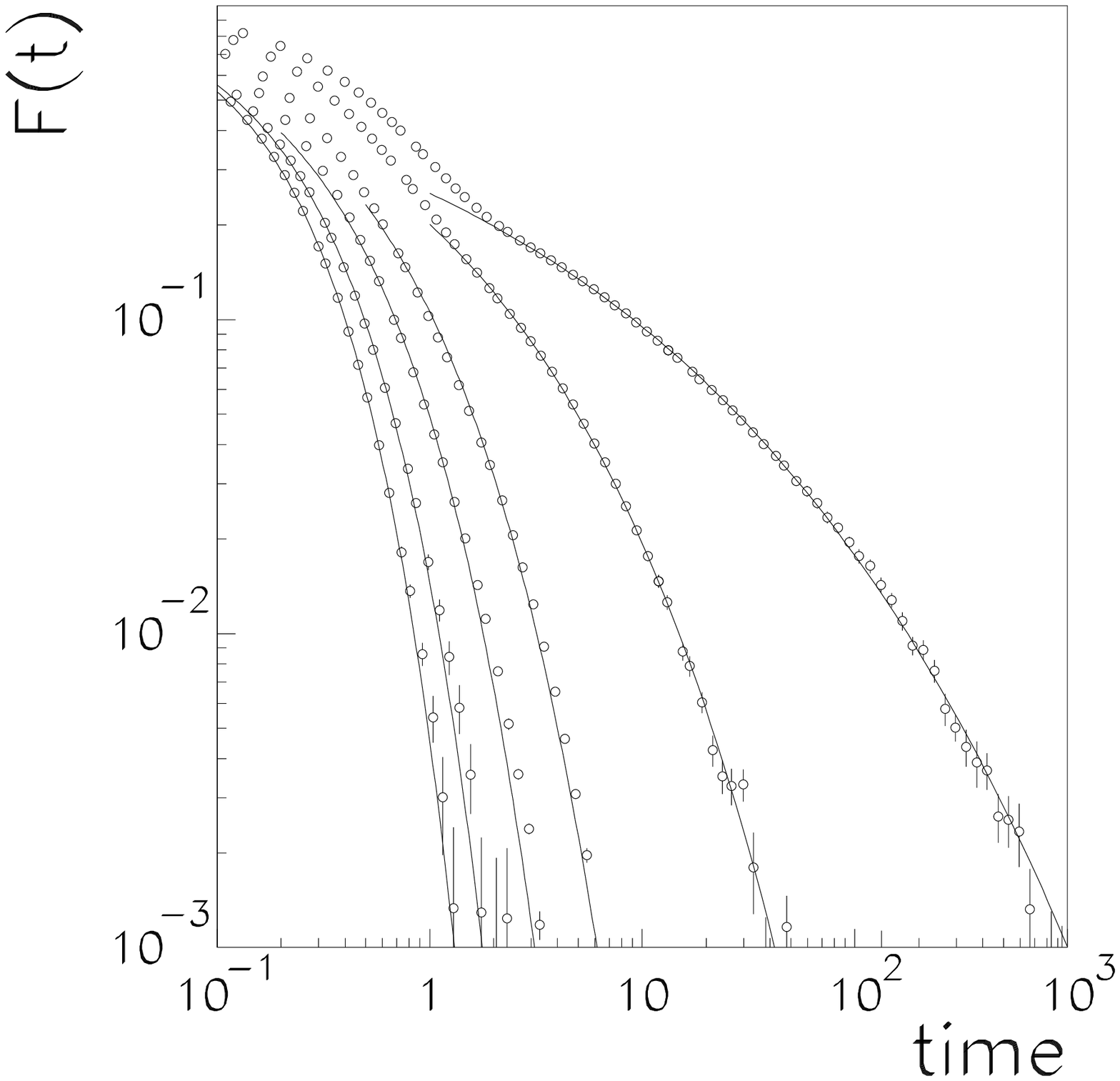}}
\end{center}
\bigskip
\begin{center}
\mbox{(b)\epsfysize=7.5cm\epsfbox{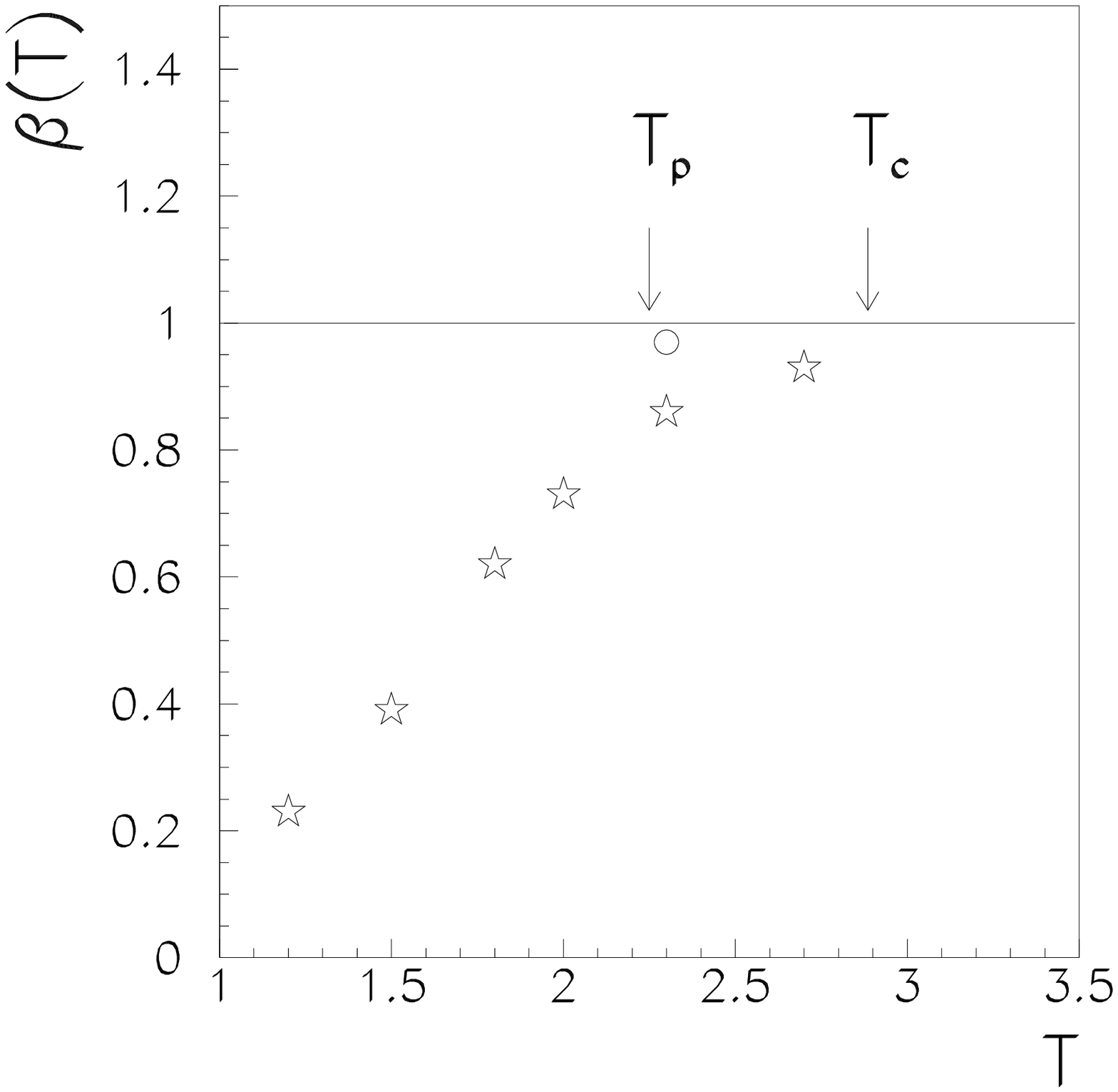}}
\end{center}
\bigskip
\caption{(a) Relaxation functions of the number of bonds in the model
with $L=32$, $q=1$, for temperatures (from left to right) 
$T=2.7$, 2.3, 2.0, 1.8, 1.5, 1.2. Solid lines are the stretched exponential
fit functions.
(b) Stretching exponent $\beta(T)$ as a function of temperature
for $L=32$ (stars). The open circle represents
a single simulation made for $L=100$ and $T=2.3$. Arrows
mark the percolation transition $T_p$ and the critic temperature of the
pure model $T_c$.}
\label{fig_relax_1}
\end{figure}

\begin{figure}
\begin{center}
\mbox{(a)\epsfysize=7.5cm\epsfbox{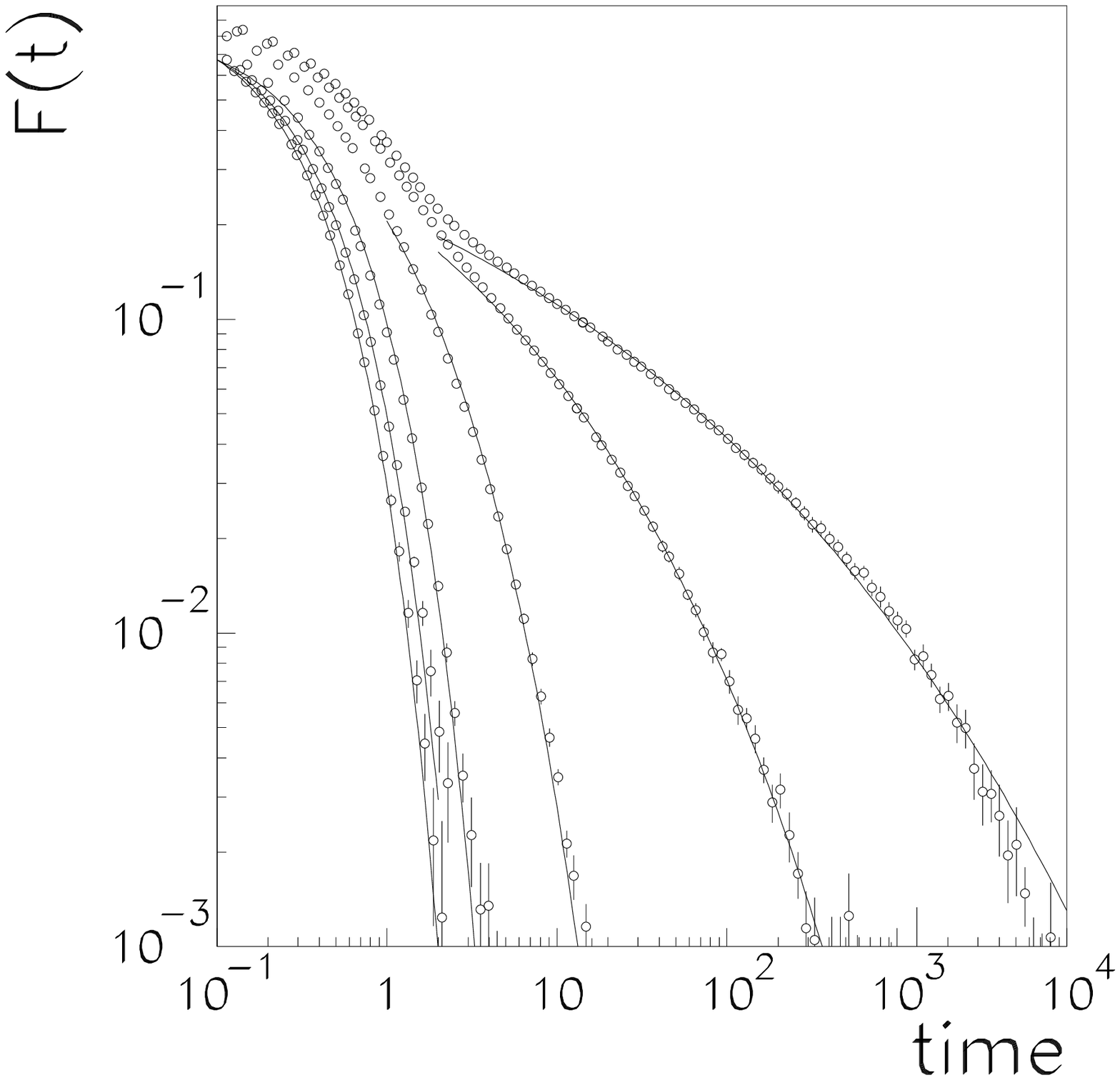}}
\end{center}
\bigskip
\begin{center}
\mbox{(b)\epsfysize=7.5cm\epsfbox{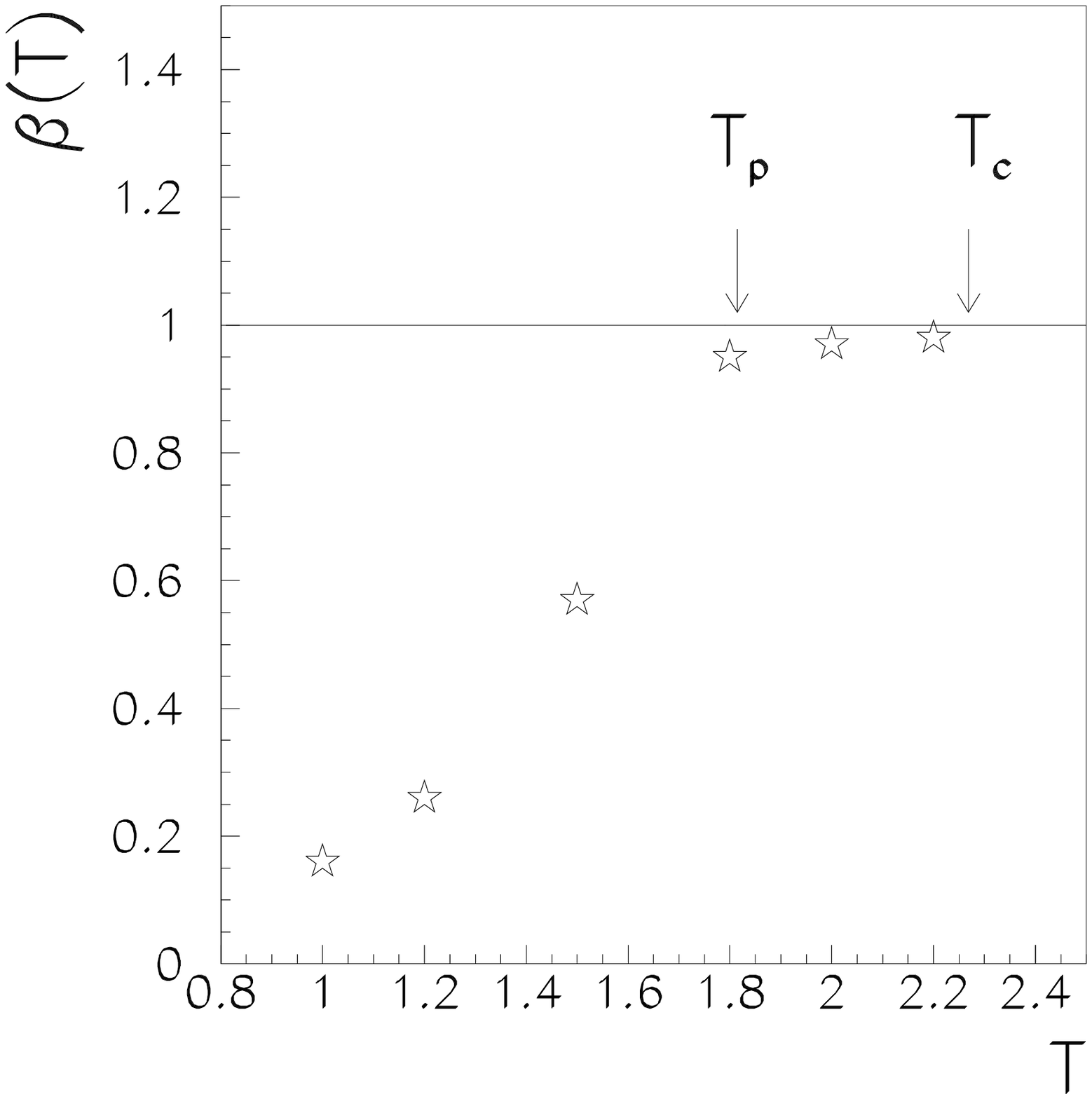}}
\end{center}
\bigskip
\caption{(a) Relaxation functions of the number of bonds in the model
with $L=32$, $q=2$, for temperatures (from left to right)
$T=2.2$, 2.0, 1.8, 1.5, 1.2, 1.0. Solid lines are the stretched exponential
fit functions.
(b) Stretching exponent $\beta(T)$ as a function of temperature.}
\label{fig_relax_2}
\end{figure}

\begin{figure}
\begin{center}
\mbox{(a)\epsfysize=7.5cm\epsfbox{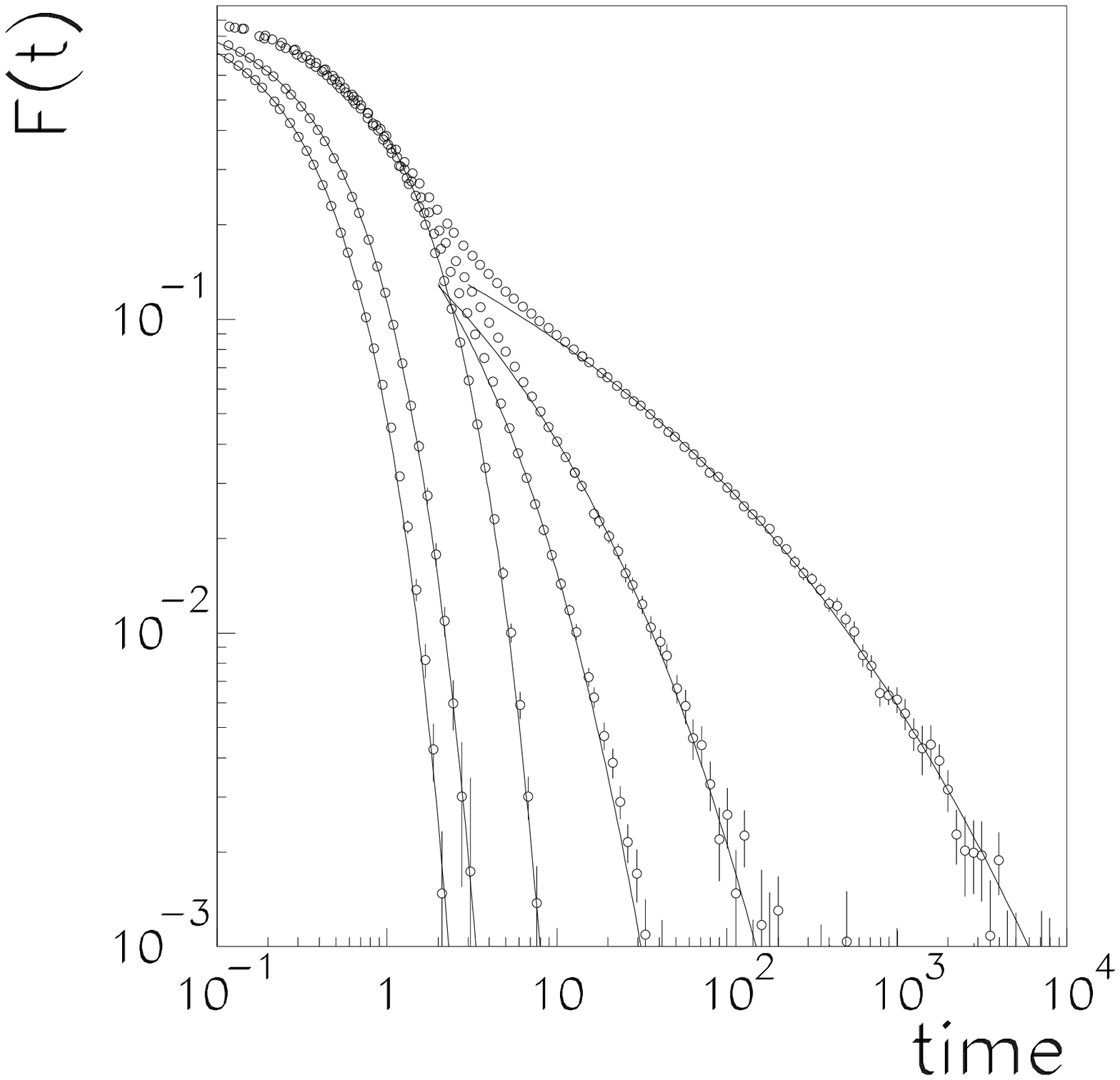}}
\end{center}
\bigskip
\begin{center}
\mbox{(b)\epsfysize=7.5cm\epsfbox{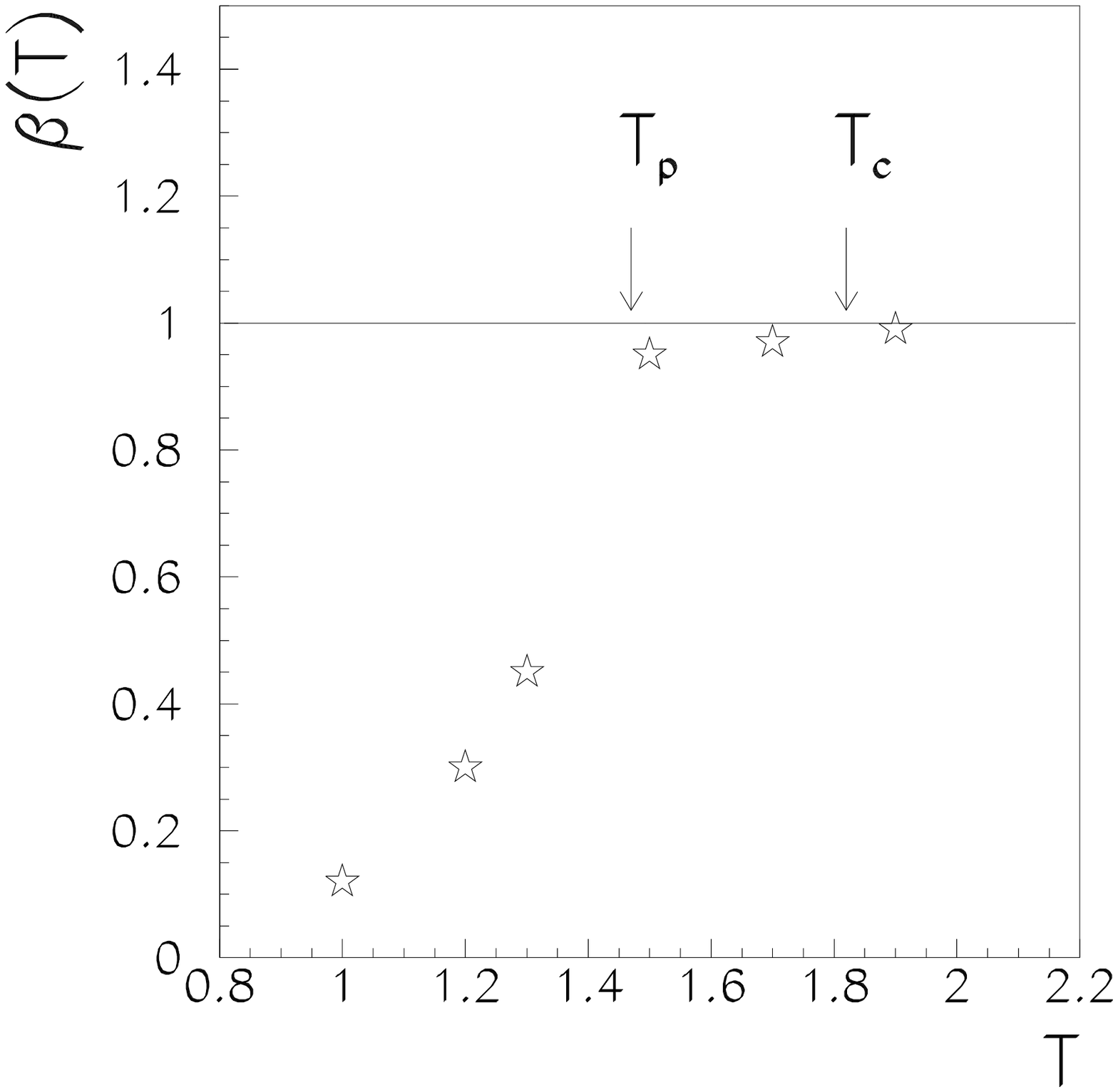}}
\end{center}
\bigskip
\caption{(a) Relaxation functions of the number of bonds in the model
with $L=32$, $q=4$, for temperatures (from left to right) 
$T=1.9$, 1.7, 1.5, 1.3, 1.2, 1.0. Solid lines are the stretched exponential
fit functions.
(b) Stretching exponent $\beta(T)$ as a function of temperature.}
\label{fig_relax_4}
\end{figure}

In Fig. \ref{fig_relax_1}(a), \ref{fig_relax_2}(a), 
and \ref{fig_relax_4}(a), we show respectively the results for 
$q=1$, 2, and 4. For all the three values of the multiplicity $q$ 
considered, we observe the same behavior. 
For high temperatures, the functions show a single exponential decay,
while at lower temperatures
they show instead a two step decay, reminiscent of what one observes
in glass forming liquids at low temperature.
This behavior can be explained by the presence, at temperatures below the
percolation transition, of a rough landscape of the free energy in
configuration space, with many minima separated by high barriers.
The short time decay
corresponds to the relaxation inside the single valley, while the
long time tail is due to the tunneling through barriers and final decay
to equilibrium ($\alpha$ relaxation).
Furthermore, the second step of relaxation functions is well fitted
by a stretched exponential function
\begin{equation}
F(t)\propto\exp -\left(\frac{t}{\tau}\right)^\beta,
\label{eq_stretch}
\end{equation}
where $\beta$ is an exponent lower than one.

In disordered and frustrated spin systems like spin glasses,
simulated by conventional spin flip,
the appearing of non exponential relaxation is believed to be caused by
the existence of unfrustrated ferromagnetic clusters
of interactions, see Randeria
{\em et al.} \cite{ref_randeria}. 
Below the ferromagnetic transition temperature $T_c$ of the pure model,
each unfrustrated cluster relaxes with a time that depends from 
its size. Due to the disorder of the interactions, 
the sizes of the unfrustrated
clusters are distributed in a wide range, giving rise to
a wide distribution of relaxation times in the model.
Therefore, according to this picture,
the temperature $T_c$ of the ferromagnetic transition of the pure model
marks the onset of non exponential relaxation.

The $q$-state frustrated percolation model is equivalent, for what concerns 
static properties, to a disordered spin system. In particular there
are unfrustrated clusters of interactions with different sizes,
due to the disorder of the variables $\epsilon_{ij}$.
On the other hand, the bond flip dynamics does not suffer from
critical slowing down near the ferromagnetic critical point 
\cite{ref_sweeny}, so we expect that the Randeria mechanism does not
apply in this case.
We have verified this point plotting the stretching exponent
$\beta(T)$ as a function
of the temperature $T$, for $q=1$, 2, and 4, as shown in
Fig. \ref{fig_relax_1}(b), \ref{fig_relax_2}(b),
and \ref{fig_relax_4}(b), respectively.
In the case of $q=2$ and 4,
it is quite evident from the data that the transition point of the
pure model $T_c$ does not mark any change in the behavior of 
relaxation functions. Instead the temperature $T_p$, at which 
clusters of bond percolate, appears as the point that marks the
onset of stretched exponential relaxation.
The case $q=1$ was not so evident from our results. 
For this reason we made 
a single simulation for a much larger system ($L=100$), at a temperature
$T=2.3$ slightly higher than the percolation threshold ($T_p=2.25$).
By fitting the function for times greater than $1.5$ MCS, where
$F(t)<0.05$, we obtain a stretching exponent $\beta=0.97$, definitely
higher than that obtained at the same temperature for $L=32$ 
($\beta=0.86$). Thus it seems  that for $q=1$ finite size effects 
at the percolation transition are more important, but nevertheless
the relaxation is asyntotically purely exponential for $T>T_p$.
In conclusion our results show that, for all the values of $q$ studied, 
in this model the relevant mechanism for the
appearing of non exponential relaxation is the percolation transition.
\section{Dynamical properties at low temperature}
\label{sec_dynamics}
\begin{figure}
\begin{center}
\mbox{(a)\epsfysize=7.5cm\epsfbox{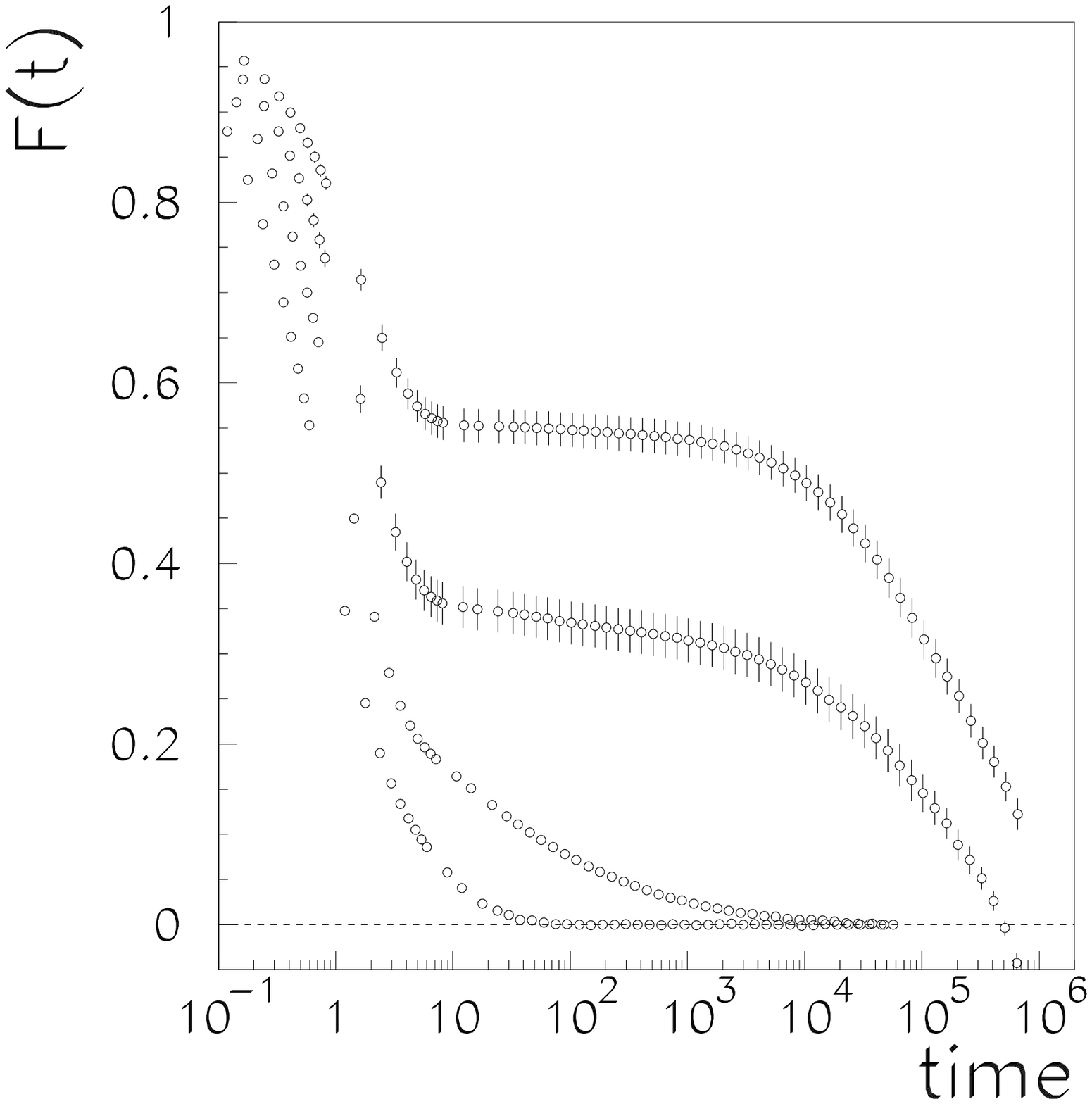}}
\end{center}
\bigskip
\begin{center}
\mbox{(b)\epsfysize=7.5cm\epsfbox{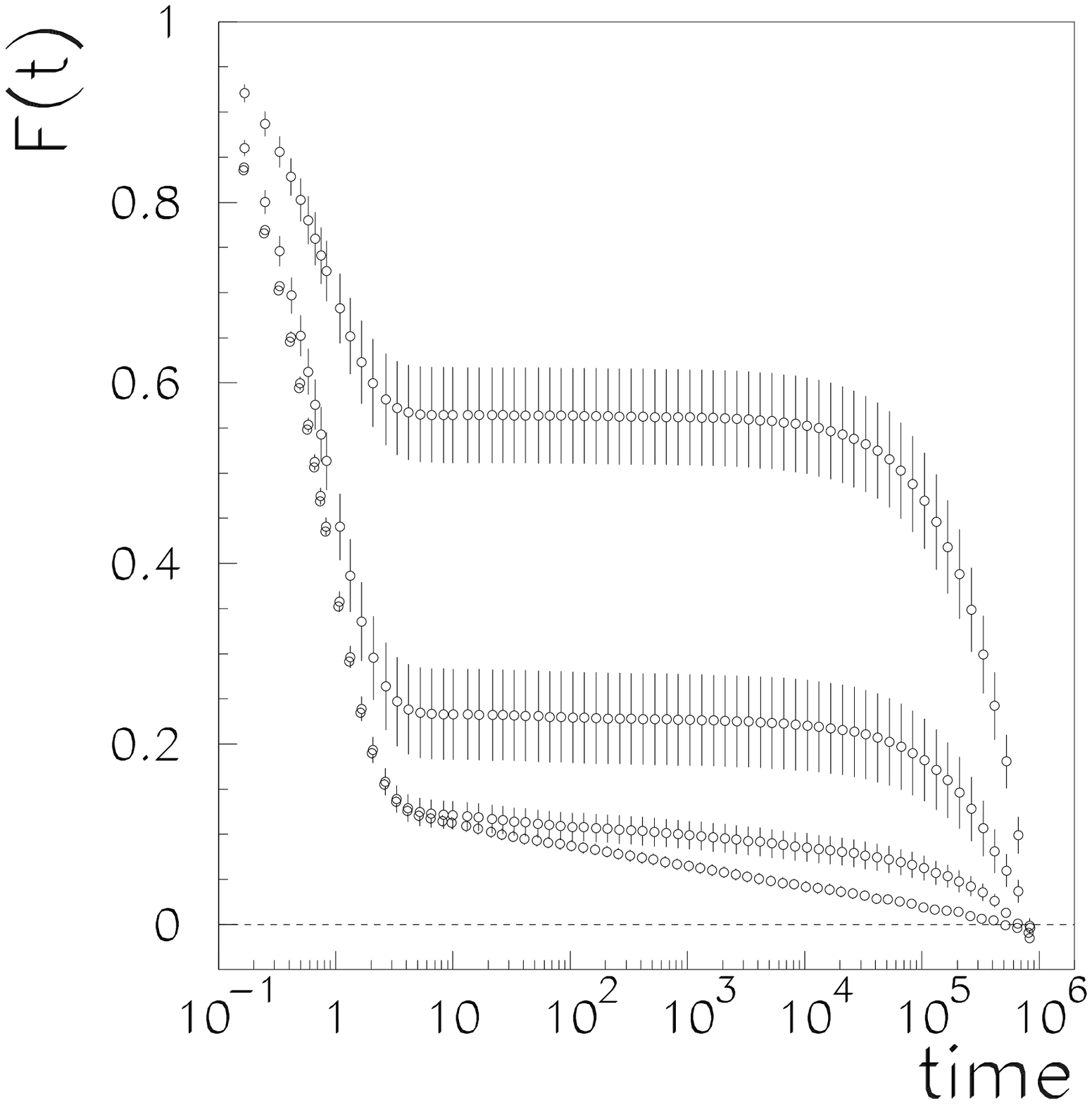}}
\end{center}
\bigskip
\caption{(a) Relaxation functions of the number of bonds in the model with
$L=40$, $q=1$, for temperatures (from bottom to top)
$T=1.5$, 1.0, 0.6, 0.5.
(b) Model with $L=40$, $q=2$, for temperatures (from bottom to top)
$T=0.6$, 0.5, 0.4, 0.35.}
\label{fig_glass}
\end{figure}

We have evaluated the relaxation functions, in the model with $q=1$ and 2,
for very low temperatures.
In Fig. \ref{fig_glass}(a) and (b) 
we show the results for $L=40$, $q=1$ and 2,
and for different temperatures, averaged
over 16 different configurations of interactions.
For the lowest temperatures, the functions do not relax
smoothly to zero. This is clearly an effect due to the
relaxation time being greater than
the total time of the run, that was between $2\times 10^6$
and $5\times 10^6$ MCS.
For such very low temperatures, the relaxation functions
show a behavior very similar to what can be observed in glass forming
liquids near or below the mode coupling theory transition temperature
\cite{ref_mct}.
A first short time decay is followed by a very long {\em plateau},
and eventually there is a final relaxation to equilibrium, for very long
times.

\begin{figure}
\begin{center}
\mbox{\epsfysize=7.5cm\epsfbox{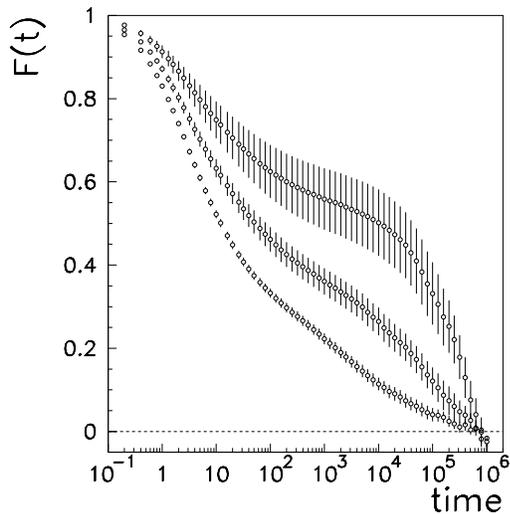}}
\end{center}
\bigskip
\caption{Relaxation functions of the energy in the spin glass
model with $L=40$, simulated by spin flip, for temperatures
(from bottom to top) $T=0.6$, 0.5, 0.4.}
\label{fig_ising}
\end{figure}

We compare this behavior to what one observes in the Ising spin glass
simulated by conventional spin flip. In Fig. \ref{fig_ising} the relaxation
functions of the energy, for the Ising spin glass on a two-dimensional
square lattice with $L=40$, and for the same temperatures of 
Fig. \ref{fig_glass}(b), are shown. Note that in this case
there is not a clear separation between the first short time decay
and the long time  tail of the functions, and the 
functions do not show any {\em plateau}.
\section{Conclusions}
We have studied the $q$-state frustrated percolation model, by means of a
``bond flip'' Monte Carlo dynamics. The model is equivalent from the
thermodynamic point of view to the Potts spin glass model (\ref{eq_potts}),
which for $q=2$ coincides with the Ising spin glass.
We have studied the percolation transition, that 
happens at a temperature $T_p$ greater than the spin glass temperature 
$T_{SG}$, and found that it belongs to the universality class of the
$q/2$-state ferromagnetic Potts model.

We have then studied the dynamical properties of the model. Differently
from what happens in spin glass systems, simulated by conventional spin 
flip, the transition temperature $T_c$ of the pure model does not play
here any role in determining the dynamical behavior. 
Instead, the percolation temperature $T_p$ appears to mark the onset of 
two step decay, and stretched exponentials in 
autocorrelation functions.

At very low temperatures the autocorrelation 
functions develop a long {\em plateau},
as observed in glass forming liquids, and predicted by the mode coupling
theory. This is a feature of the bond flip dynamics we have performed,
while the spin flip dynamics of the spin glass model,
that is thermodynamically equivalent to our model for $q=2$,
is very different, and does not show any {\em plateau}.

This results show that the frustrated percolation model bridges
the spin glass to the glass forming liquids. On one hand it is equivalent
thermodynamically (for $q=2$) to the Ising spin glass. On the other hand
the model describes a bond packing problem, where the bond are subjected
to the constraint that no frustrated loop can be closed. 
This makes the dynamics similar to
that of glass forming liquids, where irregular molecules
(or groups of molecules)
move under the constraint of some kind of steric hindrance.

To make more contact with glass forming liquids, the site version
of the frustrated percolation model has been developed \cite{ref_silvia}.
In this model particles are allowed to diffuse on the lattice,
under the constraint that no frustrated loop can be fully occupied.
Consequently it is possible to calculate the mean square displacement
and the diffusion coefficient of the particles. Numerically it is found
that also these quantities reproduce qualitatively the corresponding
quantities measured in glass forming liquids.
\section*{Acknowledgments}
This work was supported in part by the European TMR Network-Fractals
c.~n.~FMRXCT980183, and by MURST (PRIN-97).
Part of the Monte Carlo calculations were performed on the
Cray T3D and T3E at CINECA.
%
%
%
%

%
%
%
%
\raggedbottom
%
%
\newpage
\vspace*{1cm}
\vspace*{1cm}
\newpage
\vspace*{3cm}
\newpage
%
%
%
%
%
\newpage
\vspace*{1cm}
\newpage
\vspace*{1cm}
\newpage
\vspace*{1cm}
\newpage
\vspace*{1cm}
\newpage
\vspace*{1cm}
%
%
%
%
\vspace*{2cm}
\begin{table}
\begin{tabular}{|c|ccc|}
 $q$   & $T_p$ & $1/\nu$ & $\gamma$ \\
\hline
 $1$ &  $2.25\pm 0.03$   &  $0.56\pm 0.01$ &  $3.19\pm 0.05$ \\
 $2$ &  $1.814\pm 0.024$ &  $0.77\pm 0.02$ &  $2.28\pm 0.06$ \\
 $4$ &  $1.47\pm 0.04$   &  $0.95\pm 0.04$ &  $1.79\pm 0.08$ \\
\end{tabular}
\caption{Measured critical temperature and exponents
for the percolation transition in the $q$-state
frustrated percolation model. Temperatures are measured in unities
of $J/k_B$.}
\label{tab_critic}
\end{table}
\begin{table}
\begin{tabular}{|c|cccc|}
$q$ & $T_c$ & $1/\nu$ & $\gamma$ & $\alpha$ \\
\hline
$1/2$ &  $3.740$  &  $0.5611$  &  $3.2696$ & $-1.5645$ \\
$1$   &  $2.885$  &  $3/4$     &  $43/18$  & $-2/3$    \\
$2$   &  $2.269$  &  $1$       &  $7/4$    & (log)     \\
$4$   &  $1.820$  &  $3/2$     &  $7/6$    & $2/3$     \\
\end{tabular}
\caption{Critical temperature and exponents of the $q$-state ferromagnetic
Potts model. Temperatures are measured in unities of $J/k_B$.}
\label{tab_potts}
\end{table}
\end{document}